# Performance of Short-Commit in Extreme Database Environment

Muhammad Tayyab Shahzad

shahzadonline@gmail.com

School of Computing and Mathematical Sciences

Liverpool John Morres University.

Muhammad Rizwan

Muhammad.rizwan@uettaxila.edu.pk

Department of Computer Engineering

University of Engineering and Technology Taxila, Pakistan

**Abstract:**

**Atomic commit protocols are used where data integrity is more important than data availability. Two-Phase commit (2PC) is a standard commit protocol for commercial database management systems. To reduce certain drawbacks in 2PC protocol people have suggested different variance of this protocol. Short-Commit protocol is developed with an objective to achieve low cost transaction commitment cost with non-blocking capability. In this paper we have briefly explained short-commit protocol executing pattern. Experimental analysis and results are presented to support the claim that short-commit can work efficiently in extreme database environment.**

## 1 Introduction:

Commit Protocol ensures the transaction atomicity. To understand the role of commit protocols we consider an example of funds transfer from one account to another. We consider a transaction that transfers funds from account *A* to account *B*. This transaction consists of two sub transactions, one sub transaction debit sum *A* and second sub transaction credit sum *B*. In the process first transaction checks the availability of required funds in account *A* then debits it with desired amount. Then second transaction credits the account *B* with the amount which is debited from the account *A*.

Consider the situation if the second transaction fails before crediting the account *B*. This failure can occur due to any issue such as: site failure, communication failure etc. In this situation account *A* is debited without crediting the account *B*, this failure gives rise to errors. Hence, commit protocol are designed to ensure that account *A* is debited if and only if account *B* is credited. In case of failure of transaction before crediting account, then previous state of account *A* is restored to make the database consistent.

Global transactions may consist of multiple sub transactions that may execute on different remote sites. Commit protocol forces sub transaction to agree on a single outcome which means that a global transaction will commit if an only if all the sub transactions commit. In case if any of the sub transaction fails, the global transaction aborts and forces successfully executed (not committed) to abort and the previous state of the system is restored. Two-phase commit protocol (2PC) is considered standard and consists of two phases [2]. Many attempts are made to minimize the protocol execution cost by reducing the node communication or the disk write logging activity [2] [3]. In some cases blocking issue is addressed but it results in much higher execution cost of the protocol [9] [12].
.

**Related Work:**

The protocols in which emphasis is on reducing the commit cost are presumed commit protocol [3], presumed abort protocol [2], single phase commit protocol [12] and coordinator log [11] etc. In these protocols committing cost (in 2PC) reduced by minimizing the log writes or communication between the nodes or in some cases removing the entire phase. Reduction in committing cost on the other hand increases the blocking factor of the committing protocol.

In the database environment where site failure and communication failure is high , above mentioned protocols results in actually higher committing time period as compared to two phase committing protocol. This is due to the extra recovering procedures that each site has to undergo in order to preserve the status of the database.

In Three phase commit protocol [16] and optimistic commit protocol [17] the blocking issue is focused but these protocols produce much higher execution cost while providing some sort of non blocking capability. Backup site is employed in [9] in order to minimize the delay caused by the coordinator failure but this is not fully non blocking commit protocol. Extra communication between the backup site coordinator actually results in increase of committing cost of the transaction even in the absence of coordinator failure.

In new commit protocol (short commit) Non-Blocking capability is achieved having low committing cost. Extra site is employed called mediator. Mediator works parallel with coordinator. In case of coordinator failure, mediator takes the role of coordinator. This shift of responsibility is carried out without extra delay. Furthermore this protocol works equally well in the reliable environment where site and communication failures are exceptions.

## 2     Two Phase Commit Protocol:

Two phase protocol consists of two phase as name indicates. First phase is called prepared phase in which coordinator asks sites to send commit or abort vote for the transaction which has been executed (not committed). Participants log their votes before sending to the coordinator [1]. Decision phase is the second phase of the commit protocol. In decision phase if coordinator receives the commit vote or yes vote from all the participants sites then it logs the commit decision and then sends decision to all the participants. In case if it receives Abort vote or No vote from any of the participants then it sends abort decision to all the participating sites. Prepared participant after getting decision form the coordinator logs it and release the data resources pertaining to the transaction [5].

## 3     Variances of Two-phase commit protocol:

In two phase commit protocol information about committed or aborted transaction is explicitly recorded and missing information has no meaning. Presumption gives the meaning of the missing information [10] [15]. In presumed commit, information about committed transaction is not logged in to disk space which saves log writes for committed transactions. In presumed commit it is cheaper to commit a transaction than to abort [15]. Early prepare protocol gives low committing cost on the assumption that every site goes to the prepared state after acknowledging the last executed operation [13]. So there is not need for coordinator to send explicit prepared request to all participants. In Coordinator Log, logging of all participants is central on the coordinator which eliminates the need for each site to log. Central logging makes participants fully dependent on the coordinator for the recovery [11]. Further more there are very strong assumptions associated with these protocols [14] [8].

Protocols discussed up till now are the blocking protocols, it means that on the coordinator failure, prepared participant has no choice but to wait for the coordinator to recover and send decision back to prepared participant. Failure could be long and can force prepared participant to be in wait state holding the data resources, it creates the blocking state. Three phase commit protocol is the first atomic commit protocol in which blocking issue is addressed where non-blocking is achieved by adding an extra phase called pre-commit phase [16]. Optimistic commit protocol reduces blocking time period on the assumption that every transaction will commit eventually [17]. By this assumption it lets waiting transaction to access the uncommitted data of the executing transaction which contradicts the isolation property of the transaction. Backup site is used to prevent the blocking situation in one of the method but is not effective in every blocking situation. Further more commit protocol execution cost increases due to the extra communication to the backup site [9]. Mobile Commit protocols are used in many applications such as mobile banking, traffic status, and Weather information as well as many ecommerce applications [20]. These are specifically designed to accommodate in the mobile wireless environment where failure rate is high as compared to fixed line network [6] [7]. Some adopt the optimistic concurrent control strategy which does not require the locking mechanism for concurrency control [18] [19].

## 4    Short-Commit Protocol:

The main hurdle in the implementation of non-blocking commit protocol is the increased "Transaction commitment cost" which is the logging and communication cost in absence of failure. The non-blocking protocols are developed to handle a blocking situation or the coordinator failure within a certain time period but with much high transaction commitment cost. There is a need to develop a non- blocking commit protocol in which not only transaction commitment cost remains same or ideally less than the Two Phase Commit (2PC) protocol which also provides the non-blocking capability.

New site called the mediator is employed for non-blocking purpose. Mediator works as a coordinator in the background and in case if the coordinator fails in the decision phase, the mediator takes the responsibilities of the coordinator and resumes the commit process from where the coordinator failed. When the coordinator recovers from failure it only inquires one of the participants about the status of the transaction before the failure.

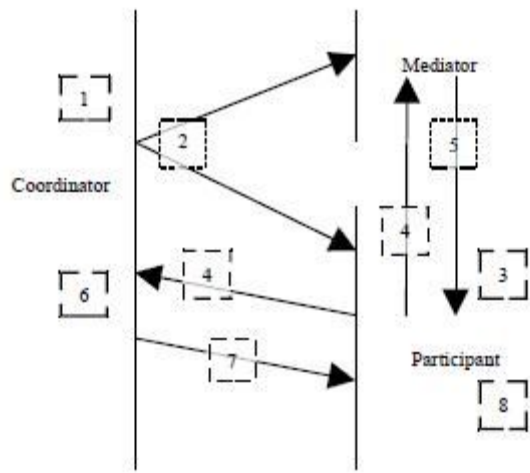

*Figure 1:Short-Commit Execution Model*

Due to the mediator involvement in the commit process participants carry out their normal operations without any extra waiting time period, even with a failed coordinator. It helps in a faster release of resources, which are held during transaction execution. This reduces the waiting time period for transactions waiting in queue to get access of resources in use. As a result the numbers of transactions per unit time increase. This feature not only makes this protocol favourable in a blocking situation but also in a reliable environment in which the data resources are shared by many processes. The Protocol is designed in a manner such that there is no need of synchronization between coordinator and the mediator which eliminates the unnecessary communication which is a basic draw back of a backup commit protocol [9].

## 4 Protocol execution:

In the following section detail algorithm of the execution of the short-commit protocol is given. Remote node communication, disk log write activity and process of recording different activity is described in detail.

### 4.1 Algorithm at Coordinator:

In STEP C1 coordinator forces a write transaction initiation record in its stable storage space and then it sends prepare message to the mediator and to all participants. At this stage coordinator waits for prepared votes from all participants after sending the prepare request to all participants. In case of any missing vote from any of the participants coordinator sends still-waiting message or second prepare request to all participants and goes to the wait state again.

```
STEP C1            Prepare Request
    {
    Write      Transaction initiation record in Stable storage space
    Send       Prepare message to mediator and all Participants
    Step C1a   Do while    Votes not received from all participants
    Wait
    On Timeout
    Send  Still-waiting message to all participants
    Go To Step C1a
    End Do
    Go To STEP C2      Decision   }
```

*Figure 2: Prepare Request at Coordinator*

After getting commit votes from all participants or abort decision from any of the participants then the algorithm proceeds to STEP C2.

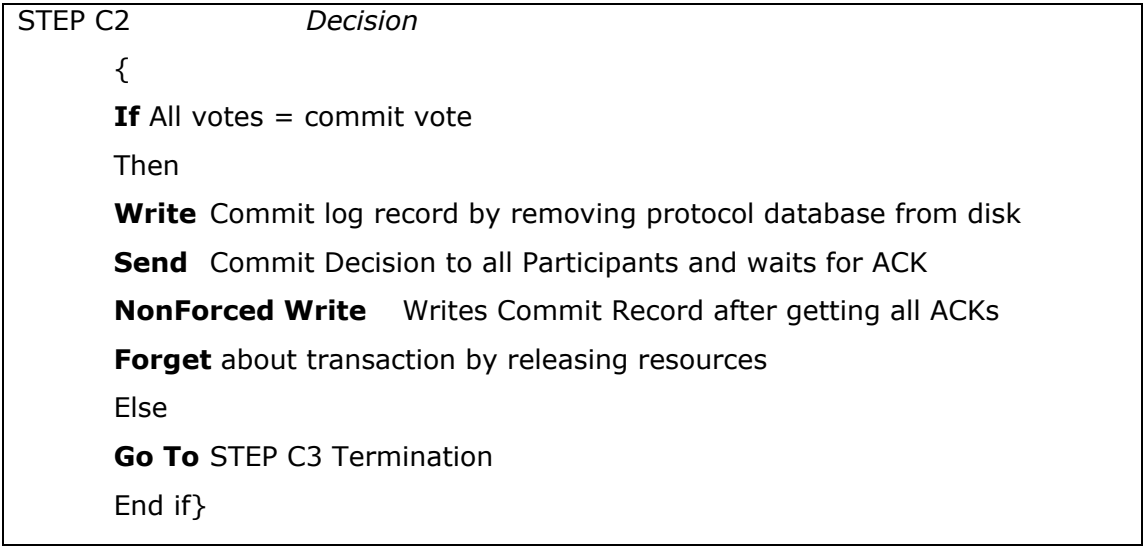

*Figure 3: Decision at Coordinator*

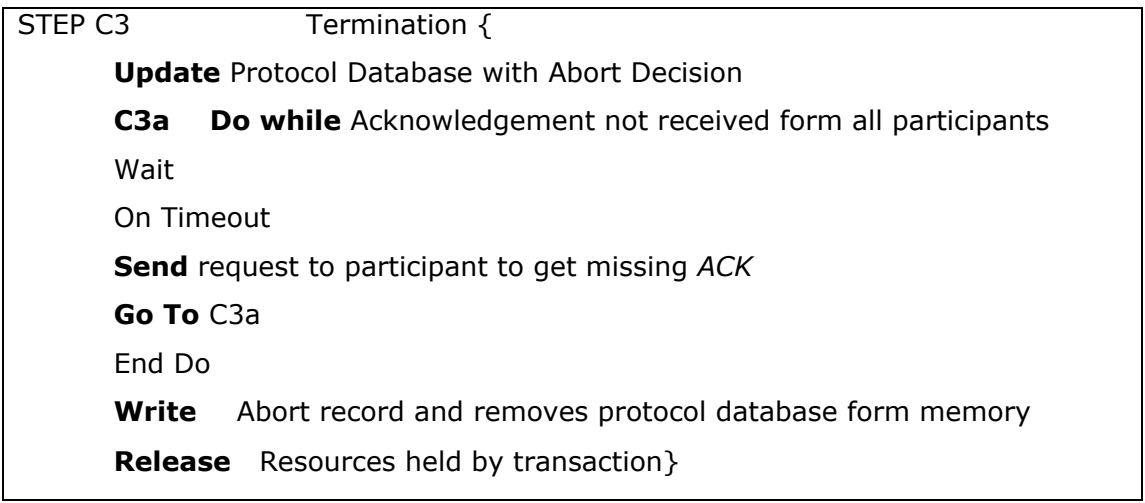

*Figure 4: Termination at Coordinator*

If the coordinator receives a *commit* vote from each of the participants then it forces a *write* commit log record into the stable storage space. This logically removes the entry from protocol database and sends commit decision to all participants. In case of any of abort decision from any of the participants STEP C3 of termination starts.

STEP C3 starts when coordinator gets abort decision from any of the participants as reply of prepare request. On getting abort decision the coordinator force writes abort log record and waits for the acknowledgement from other prepared participants. After getting all the acknowledgements from prepared participants, coordinator releases the resources pertaining to the transaction.

## 4.2 Algorithm at Mediator

```
Begin (After getting Prepare message from Coordinator)
STEP M1      Building protocol database {
             Build protocol database in main memory
             Step M1a    Do while    Votes not received from all participants
             Wait   On Timeout
             Send   Still-waiting message to all participants
             Go To Step M1a
             End Do
             Go To STEP M2 Decision}
```

*Figure 5: Building Protocol Database at Mediator*

```
STEP M2      Decision
             {
             If All votes = commit vote
             Then
             Send   Commit Decision to all Participants and waits for ACKs
             Forget about transaction after getting ACKs from all participants
             Else
             Go To  STEP C3 Termination
             End if
             }
```

*Figure 6: Decision at Mediator*

The mediator builds its protocol database in its main memory after it gets the prepared message from the coordinator. Mediator waits for votes from all participants. After it gets commit votes from all of the participants or abort decision from any of the participant it goes to STEP M2 Decision.

On receiving commit votes from all participants, Mediator sends commit decision to all prepared participants and waits for the acknowledgment of the decision. In case of abort decision from any of the participant mediator writes abort decision in its protocol database.

# 5      Simulation:

In this section we evaluate the protocol performance by conducting simulation study for three atomic commit protocols. By changing the different performance factors we have compared the 2PC protocol Presumed Commit and Short-commit protocol. Simulation is developed in JAVA programming language. There are 15 sites are used as cohorts from which 5 sites are chosen randomly for transaction execution including coordinator and mediator. There are 2500 memory locations or data pages on each site. 5 data pages are accessed by each transaction which is randomly chosen. Delays are introduced to simulate the delay associated with the forced log write activity and communication delay.

In simulation we have checked the performance differences of different protocols in diverse database environments. Multiprogramming level (MPL) of each site and Failure probability has strong impact on the performance of the commit protocol. Performance of protocol increases by increasing the MPL level of the transaction. MPL level of the particular site is the number of transactions which can execute simultaneously. By increasing MPL of the site, performance of protocol increases to a certain level, after increase of certain level higher values can produce data resource contentions and there would be more dead locks and blocking situations in which a transaction has to wait to executing transaction to complete. Level where data contentions start depends on the availability of resources.

Site failures also have strong impact on the performance of the protocol. Site failures results in many transactions to abort or could create blocking situations if the failed site is coordinator. As result of site failures, not only probability that a transaction will abort eventually increases but it also increases the protocol execution time period.

## 5.1 Base line Experiment Values:

In our simulation we have defined some base line values for some parameters. By using specific values, this ensures considerable difference between the protocols in terms of performance.

Table 1: Base Line Experiment Values

| NumSites | 15 | DBSize | 2500 pages |
|---|---|---|---|
| TransType | Sequential | DistDegree | 4 |
| CohortSize | 5 pages | MPL | 4 – 8 |
| NumCPUs | 1 | NumDataDisks | 2 |
| NumLogDisk | 1 | PageCPU | 5ms |
| PageDisk | 15ms | MsgDelay | 50ms |

*NumSites:* Are the total number of sites chosen randomly.

*DBSize:* Is the number of database pages. Locks are placed on page level.

*TransType:* On each site the transaction executes in a sequential fashion.

*DistDegree:* Randomly chosen sites for transaction execution.

*CohortSize:* Number of Data pages accessed by transaction.

*NumDataDisk:* There are 2 disks to store actual data in the database.

*NumLogDisk:* Disk used to record the log for the execution of commit protocol.

*NumPagDisk:* Time needed for each write operation on to the disk space.

*PageCPU:* Time consumed by CPU for each write operation, is 5 milliseconds.

MsgDelay: Is the propagation delay on the network.

## 5.2     Experiments and Results:

In this section we have described two different simulation studies. In first we have varied the MPL level of the site and analyzed its impact on the performance of different commit protocols. Increasing the MPL level can increases system throughput in terms of increase in the number of executing transaction per unit time. In second section we have analysed the impact of different failure frequency on the commit protocols. Failure of sites could result in increase of number of aborted transactions.

### 5.2.1 Multiprogramming level (MPL):

Each site has its specific multiprogramming level; any transaction which violates the MPL limit is aborted. The MPL limit restricts the number of executing transactions at one time on each site. The particular value of the MPL limit is chosen on each site to maintain resources and keep data contentions at a low level. In this experiment we have executed 20,000 transactions with failure probability of 0.005. The MPL value at each site is changed from 4 to 8 to analyze the behaviour of each protocol.

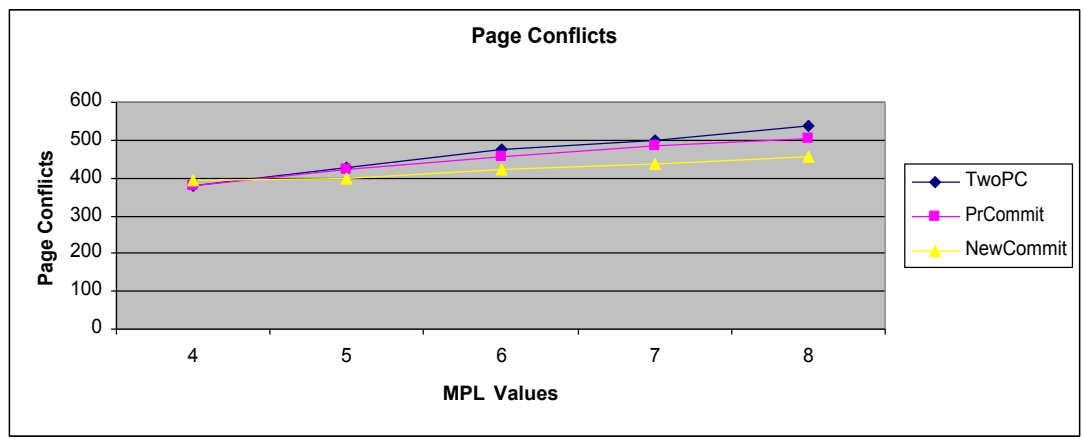

Figure 7: Page Conflicts Chart

Page conflict occurs when executing transaction tries to access the page which is already in use of the other transaction. Transaction aborts after discovering the page conflict.

In Fig.8 all protocols have same values of page conflicts initially because not many transactions enter in to the system. When MPL level of each site increases then comparatively more transaction enters in to the system. Probability that a page conflicts occur will increase with the increase in the MPL level. The protocol has shorter execution time causes the transaction finish sooner. As a result time to hold the data resources decreases causing the page conflicts to decrease. Short-Commit protocol has lower page conflicts due to its short executing time period as compared to 2PC and presumed Commit protocol.

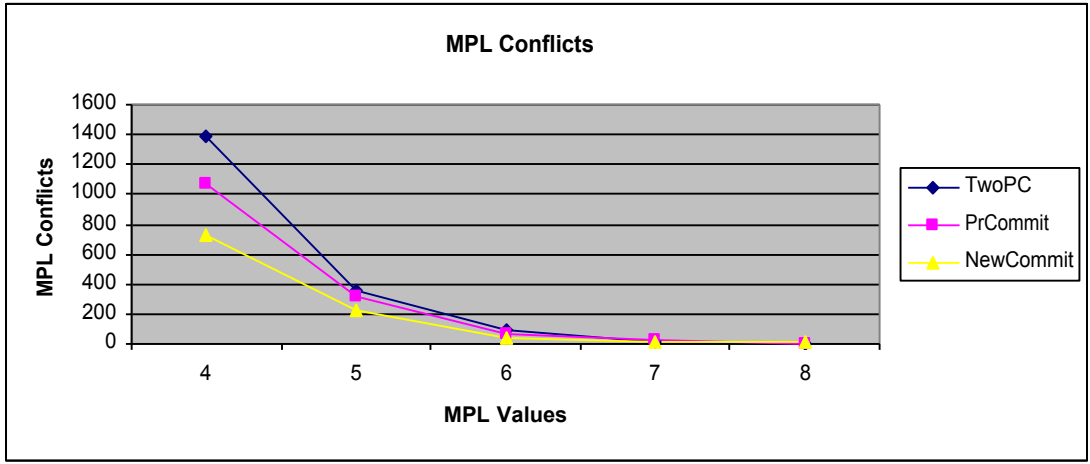

Figure 8: M PL Conflicts Chart

MPL conflicts occur when amount of transaction exceeds than the MPL level of the site. It is obvious that by increasing the MPL level the MPL level conflicts will increase. Transaction in short-commit protocol leaves the system early because of its short executing time period as compared to 2PC and presumed commit. With *MPL* value set at 4 there is significant difference in protocol *MPL* conflicts values. 2PC has the highest ratio and New Commit has the lowest ratio of Page conflicts. As *MPL* level for each site increase, causes *MPL* conflicts for all for protocols to decrease rapidly as shown in Figure 8. At *MPL* value 7 and 8 all protocols have very low *MPL* conflicts ratio. However increase in *MPL* values from here causes an increase in the page conflicts because the amount of transactions executing in the system increases and chance of accessing the same data resources which results in increase in values of data contentions.

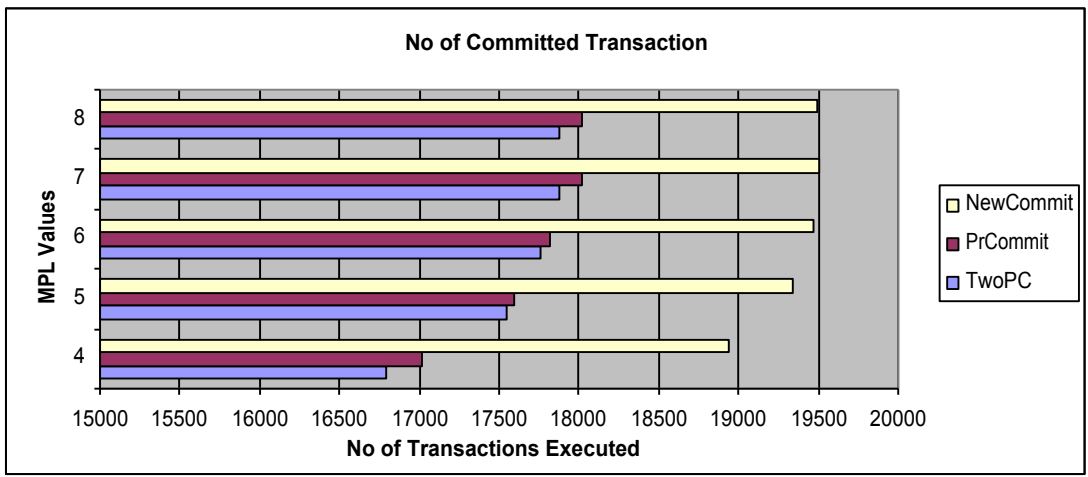

Figure 9: No. of Committed Transaction

Resource contentions of all three commit protocols are affected by changing MPL levels of sites. Average commit time, average abort time and *MPL* level have impact on the performance of the protocol. This performance difference becomes prominent when we analyze the number of committed and the number of aborted transactions for high *MPL* value as shown in Figure.9

**5.2.2  Impact of Failures:**

Blocking is one of the main drawback of many cost effective protocol. Blocking occurs when coordinator fails before sending the decision to the prepared participants. In blocking state prepared participant has to wait holding the data resources locks until coordinator recovers from failure and sends decision to the waiting site. One of the main features of the New Commit protocol is that it is resilient to site failures as compared to other protocols.

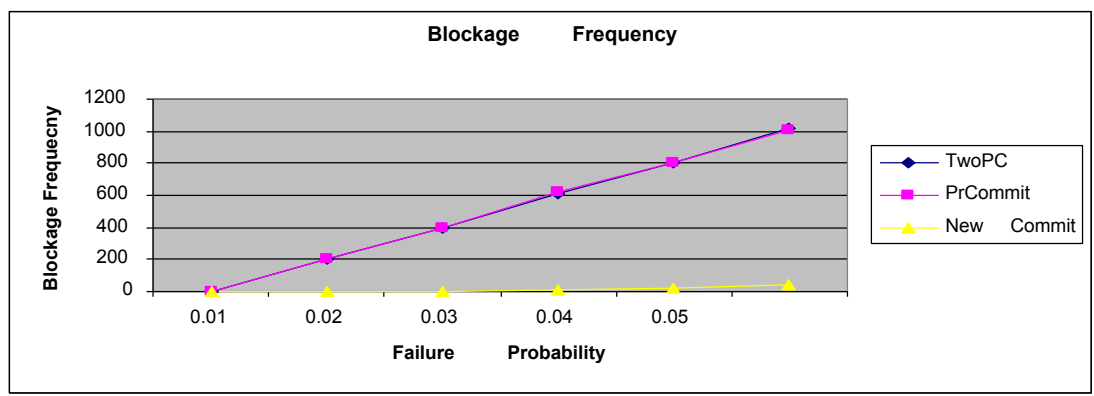

Figure 10: Blockage Frequencies

2PC and presumed commit has same failure handling procedures as shown in Fig.10. they have almost same blocking frequency. Short-commit performs much better having very low blocking frequency because of its strong failure handling procedures. There are different types of failures.

- *If coordinator fails before sending decision:* in 2PC and presumed commit protocol participant has to wait until coordinator recovers and sends decision to the prepared participants. In short-commit the prepared participant issues the abort decision if it time stamp expires in waiting of the commit decision from the coordinator. As shown Fig.10 short-commit has very low blocking frequency because in short-commit blocking will only credit if and only if both coordinator and mediator fails.

- *If coordinator fails after sending* situation *decision:* this situation does not create any impact on the blocking situation as decision has already been issued. Prepared participant will release the resources after getting decision.
- *Participant failure:* In 2PC and presumed commit, coordinator waits participant to recover and send vote until its time stamp expires. In the absence of vote from any of the participants coordinator sends abort decision to every prepared participant. In short commit protocol failed participants only delays the commit decision from coordinator and mediator because coordinator sends second prepare message instead of sending abort decision. Due to this delay any prepared participant may decide to abort in case if its time stamp expires. Aborting participant will sends abort decision directly to every prepared participant as well as to the coordinator and mediator. This helps to reduce the protocol execution time period for the aborting transaction.

Figure 11, shows the number of aborted transactions for different failure values. New Commit has the lowest aborted percentage among these protocols due to the factors explained above. For 2PC and Presumed Commit, the number of aborted transactions increases as we increase the failure probability because of missing prepared votes due to the failure of cohorts. In the new commit protocol the aborted transactions are due to the page conflicts and M PL conflicts.

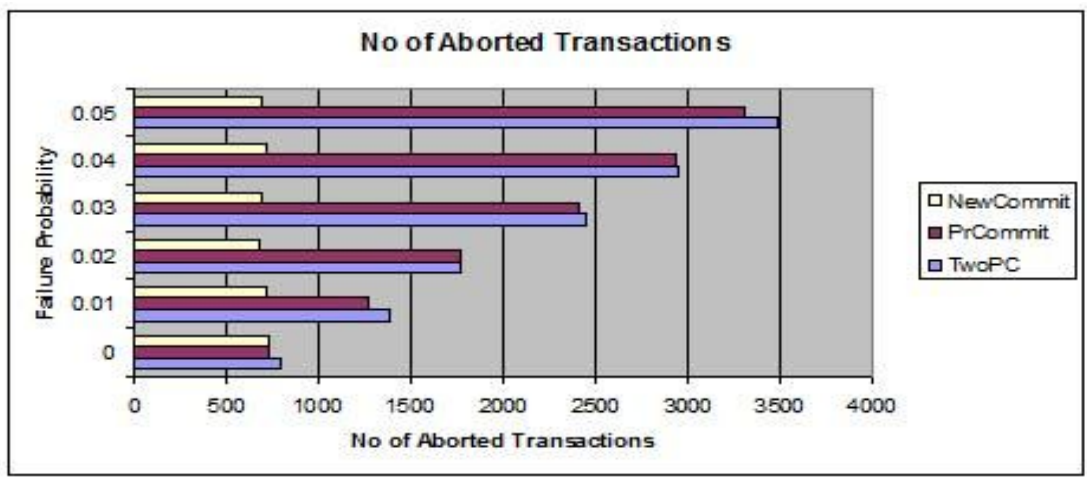

*Figure 11: No. of Aborted Transactions*

# 6     Conclusion:

Protocol having shorter commit time as compared to other commit protocols performs better in the environment where there is huge transaction load. Short-Commit protocol works comparatively well in the environment where numbers of transaction are huge because of low execution time. High failure rates and blocking factor increases the probability of transactions to get aborted. It also increases the protocol execution time period. Short-Commit protocol is non blocking protocol have much better failure handling procedures as compared to other commit protocols which helps in achieving high system throughput.

By changing multiprogramming level of the site affects the performance of the protocol. Initially system throughput increases because more transaction enters in to the system. After certain level of multiprogramming value, performance starts to decrease because of resource and data contentions. By increasing value of MPL, page Conflicts increases however number of page conflicts in New Commit is low as compared to Two Phase Commit and Presumed Commit. Multiprogramming Conflicts  values at specific MPL level is less in New Commit Protocol as compared to other protocols. Due to better performance in MPL conflicts and Page Conflicts, number of committed transaction in Short Commit protocol is higher than other mentioned protocols.

# 7 Appendix A: Simulation Results:

In this section we have presented the statistical data which is gathered during the execution of different experiments. Multiprogramming Level of each site is varied and their effect on other characteristics of the protocol is observed. Value of *MPL* varies from 4 to 8 for each site as shown in Table 2.

*Table 2: Multiprogramming Level*

| PROTOCOL | 2PC | PrCom | NewCom | 2PC | PrCom | NewCom |
|---|---|---|---|---|---|---|
| MPLValues | 4 | 4 | 4 | 5 | 5 | 5 |
| CommittedTransactions | 16797.5 | 17015.5 | 18939.7 | 17543.7 | 17588.5 | 19339.5 |
| AvgCommitTime | 306.2 | 259 | 238 | 308.5 | 259.2 | 237.7 |
| UncertainCommitTime | 159.5 | 162.5 | 141.5 | 160.5 | 162.7 | 141 |
| AbortedTransactions | 3201.2 | 2983.2 | 1058 | 2455 | 2410.5 | 655.7 |
| AvgAbortTime | 334.2 | 343 | 167 | 344.5 | 345 | 176.5 |
| UncertainAbortTime | 185.7 | 170 | 67 | 195 | 174.2 | 76.2 |
| BlockingFrequency | 608 | 593 | 15 | 607.7 | 593.2 | 12.7 |
| CoordinatorFailure | 608 | 593 | 480.7 | 607.7 | 593.5 | 521.2 |
| MediatorFailure | 0 | 0 | 481.7 | 0 | 0 | 532.2 |
| NullTransactionRestarts | 1723.5 | 1740.5 | 1711 | 1753.5 | 1713.7 | 1785.7 |
| NullCommit | 0 | 0 | 1629.7 | 0 | 0 | 1698.5 |
| NullAbort | 1723.5 | 1740.5 | 81 | 1753.5 | 1713.7 | 96.7 |
| PageConflicts | 380 | 380 | 394 | 428.2 | 420.2 | 400.5 |
| MPLConflicts | 1384.5 | 1068.5 | 728 | 355 | 319 | 228 |
| CommitPercentage | 83.98 | 85.07 | 94.69 | 87.71 | 87.94 | 96.69 |
| AbortPercentage | 16.0 | 14.91 | 5.29 | 12.27 | 12.05 | 3.27 |
| | | | | | | |
| TotalTransactions=20000 | 2PC | PrCom | NewCom | 2PC | PrCom | NewCom |
| MPLValues | 6 | 6 | 6 | 7 | 7 | 7 |
| CommittedTransactions | 17757.7 | 17813.7 | 19472.2 | 17874.5 | 18024.2 | 19501.2 |
| AvgCommitTime | 310.2 | 259.2 | 238 | 311.7 | 258.5 | 235.7 |
| UncertainCommitTime | 162 | 163 | 141.7 | 162 | 162.2 | 140.2 |
| AbortedTransactions | 2242.2 | 2185.7 | 525.7 | 2125 | 1988.7 | 494.2 |
| AvgAbortTime | 351.5 | 353 | 181.5 | 354.7 | 351.5 | 179.2 |
| UncertainAbortTime | 200.5 | 180 | 81.2 | 203.2 | 179 | 80.2 |
| BlockingFrequency | 603.7 | 613.7 | 14.5 | 600.7 | 593.2 | 16.7 |
| CoordinatorFailure | 603.7 | 613.7 | 517.2 | 600.7 | 593.2 | 539 |
| MediatorFailure | 0 | 0 | 531.5 | 0 | 0 | 506 |
| NullTransactionRestarts | 1762.7 | 1711.7 | 1775.2 | 1725 | 1762 | 1753 |
| NullCommit | 0 | 0 | 1680.7 | 0 | 0 | 1666 |
| NullAbort | 1762.7 | 1711.7 | 94.2 | 1737 | 1762 | 87 |
| PageConflicts | 476.7 | 457.2 | 421.7 | 501 | 483 | 436 |
| MPLConflicts | 86 | 67.2 | 42 | 14.2 | 21.2 | 13.2 |
| CommitPercentage | 88.78 | 89.06 | 97.36 | 89.37 | 90.12 | 97.50 |
| AbortPercentage | 11.21 | 10.92 | 2.62 | 10.62 | 9.94 | 2.47 |

| PROTOCOL | 2PC | PrCom | NewCom |
|---|---|---|---|
| MPLValues | 8 | 8 | 8 |
| CommittedTransactions | 17872.5 | 18013.2 | 19492.7 |
| AvgCommitTime | 304 | 259.7 | 236.7 |
| UncertainCommitTime | 157.7 | 163 | 140.7 |
| AbortedTransactions | 2126.7 | 1986.2 | 504.2 |
| AvgAbortTime | 345 | 352.5 | 182.5 |
| UncertainAbortTime | 197 | 179.5 | 82.5 |
| BlockingFrequency | 593.7 | 623.2 | 15.75 |
| CoordinatorFailure | 593.7 | 623.2 | 522.2 |
| MediatorFailure | 0 | 0 | 519.5 |
| NullTransactionRestarts | 1705 | 1750.7 | 870 |
| NullCommit | 0 | 0 | 1642.2 |
| NullAbort | 1705 | 1750.7 | 95.5 |
| PageConflicts | 536.5 | 502.2 | 453.7 |
| MPLConflicts | 6 | 2.5 | 14.5 |
| CommitPercentage | 89.36 | 90.06 | 97.46 |
| AbortPercentage | 10.63 | 9.93 | 2.52 |

In Table 3 data about different parameters of protocols is collected and presented at different failure rates. In this experiment 20,000 transactions have been executed with the failure probability range from 0 to 0.005.

*Table 3: Failure Probability*

| PROTOCOL | 2PC | PrCom | NewCom | 2PC | PrCom | NewCom |
|---|---|---|---|---|---|---|
| FailureProbability | 0 | 0 | 0 | 0.01 | 0.01 | 0.01 |
| CommittedTransactions | 19201.5 | 19257.7 | 19271.7 | 18623.5 | 18730.7 | 19274 |
| AvgCommitTime | 299.7 | 249.2 | 218.2 | 302.2 | 252 | 223.5 |
| UncertainCommitTime | 150.7 | 152.7 | 123.7 | 153.5 | 156 | 129 |
| AbortedTransactions | 798.5 | 742.2 | 737.2 | 1376.5 | 1269.2 | 723 |
| AvgAbortTime | 299.2 | 303.7 | 151.2 | 324.7 | 326.7 | 157.5 |
| UncertainAbortTime | 150 | 127 | 52.7 | 175.5 | 153.5 | 59.25 |
| BlockingFrequency | 0 | 0 | 0 | 203.2 | 199.2 | 1.5 |
| CoordinatorFailure | 0 | 0 | 0 | 203.2 | 199.2 | 190.5 |
| MediatorFailure | 0 | 0 | 0 | 0 | 0 | 182 |
| NullTransactionRestarts | 0 | 0 | 0 | 588.5 | 590.2 | 588.7 |
| NullCommit | 0 | 0 | 0 | 0 | 0 | 557.7 |
| NullAbort | 0 | 0 | 0 | 588.5 | 590.2 | 31 |
| PageConflicts | 484.2 | 509.5 | 388.5 | 479.7 | 481.2 | 406.2 |
| MPLConflicts | 370 | 248.7 | 395 | 371.5 | 253 | 342 |
| CommitPercentage | 96.00 | 96.28 | 96.35 | 93.11 | 93.65 | 96.37 |
| AbortPercentage | 3.99 | 3.71 | 3.68 | 6.88 | 6.34 | 3.61 |

| PROTOCOL | 2PC | PrCom | NewCom | 2PC | PrCom | NewCom |
|---|---|---|---|---|---|---|
| **FailureProbability** | **0.02** | **0.02** | **0.02** | **0.03** | **0.03** | **0.03** |
| **CommittedTransactions** | 16981.7 | 18228 | 19313.7 | 17550.2 | 17587.5 | 19301 |
| **AvgCommitTime** | 304.5 | 257 | 229.5 | 309.2 | 260 | 234.2 |
| **UncertainCommitTime** | 156.5 | 159.5 | 134.7 | 160.5 | 163.2 | 138 |
| **AbortedTransactions** | 1767.5 | 1772 | 679.5 | 2449.5 | 2412.5 | 698.7 |
| **AvgAbortTime** | 336.7 | 343 | 164.5 | 345.2 | 348.75 | 172 |
| **UncertainAbortTime** | 187.5 | 169 | 65.5 | 194.7 | 175.75 | 72.5 |
| **BlockingFrequency** | 396.7 | 401.7 | 4.7 | 607.2 | 621.5 | 7 |
| **CoordinatorFailure** | 362 | 401.7 | 336.2 | 599.5 | 621.5 | 423 |
| **MediatorFailure** | 0 | 0 | 358.5 | 0 | 0 | 446.2 |
| **NullTransactionRestarts** | 1093 | 1131.7 | 1197.5 | 1712 | 1742.7 | 1472.2 |
| **NullCommit** | 0 | 0 | 1142 | 0 | 0 | 1389.2 |
| **NullAbort** | 1093 | 1131.7 | 55.25 | 1712 | 1742.75 | 83 |
| **PageConflicts** | 423 | 455.5 | 407.7 | 441.7 | 455.5 | 397 |
| **MPLConflicts** | 361.5 | 242.7 | 283 | 385.2 | 285.5 | 249.2 |
| **CommitPercentage** | 84.90 | 91.14 | 96.56 | 87.75 | 87.93 | 96.50 |
| **AbortPercentage** | 8.83 | 8.86 | 3.39 | 12.24 | 12.062 | 3.49 |
| | | | | | | |
| **PROTOCOL** | **2PC** | **PrCom** | **NewCom** | **2PC** | **PrCom** | **NewCom** |
| **FailureProbability** | **0.04** | **0.04** | **0.04** | **0.05** | **0.05** | **0.05** |
| **CommittedTransactions** | 17047.7 | 17060.2 | 19283.2 | 16511.2 | 16694 | 19300 |
| **AvgCommitTime** | 314.7 | 262.5 | 241.7 | 318.5 | 266 | 247.2 |
| **UncertainCommitTime** | 165.2 | 166.2 | 145.5 | 169.2 | 170 | 151 |
| **AbortedTransactions** | 2950.7 | 2939.7 | 716.25 | 3488.7 | 3306 | 699.7 |
| **AvgAbortTime** | 356.2 | 379.2 | 178.5 | 361.2 | 359.7 | 182.7 |
| **UncertainAbortTime** | 205.7 | 182.2 | 78.5 | 210.2 | 187.7 | 82.7 |
| **BlockingFrequency** | 808.7 | 804.5 | 25.75 | 1014.5 | 1012 | 39.5 |
| **CoordinatorFailure** | 808.7 | 807 | 656.5 | 1014.5 | 1012 | 797.5 |
| **MediatorFailure** | 0 | 0 | 644.25 | 0 | 0 | 819.7 |
| **NullTransactionRestarts** | 2275.5 | 2255.2 | 2288.5 | 2851.7 | 2874.5 | 901.5 |
| **NullCommit** | 0 | 0 | 2172.2 | 0 | 0 | 2688.7 |
| **NullAbort** | 2275.5 | 2255.5 | 114.75 | 2851.7 | 2874.7 | 129.5 |
| **PageConflicts** | 440 | 447.25 | 445.25 | 436.7 | 432 | 433.5 |
| **MPLConflicts** | 344.5 | 324 | 243 | 282.5 | 360.2 | 214.2 |
| **CommitPercentage** | 85.23 | 85.30 | 96.41 | 82.55 | 83.47 | 96.5 |
| **AbortPercentage** | 14.75 | 14.69 | 3.58 | 17.44 | 16.53 | 3.49 |